\documentclass[aps,prb,reprint,superscriptaddress,showpacs]{revtex4-1}
\usepackage{graphicx}

\usepackage{color}
\usepackage{ulem}

\begin{document}

\title{Superconductivity in MgPtSi: An orthorhombic variant of MgB$_2$}

\author{Kazutaka Kudo}
\email{kudo@science.okayama-u.ac.jp}
\affiliation{Department of Physics, Okayama University, Okayama 700-8530, Japan}
\affiliation{Research Center of New Functional Materials for Energy Production, Storage, and Transport, Okayama University, Okayama 700-8530, Japan}

\author{Kazunori Fujimura}
\affiliation{Department of Physics, Okayama University, Okayama 700-8530, Japan}

\author{Seiichiro Onari}
\affiliation{Department of Physics, Okayama University, Okayama 700-8530, Japan}

\author{Hiromi Ota}
\affiliation{Division of Instrumental Analysis, Department of Instrumental Analysis and Cryogenics, Advanced Science Research Center, Okayama University, Okayama 700-8530, Japan}

\author{Minoru Nohara}
\affiliation{Department of Physics, Okayama University, Okayama 700-8530, Japan}
\affiliation{Research Center of New Functional Materials for Energy Production, Storage, and Transport, Okayama University, Okayama 700-8530, Japan}

\date{}

\begin{abstract}
A ternary compound, MgPtSi, was synthesized by solid-state reaction. 
An examination of the compound by powder X-ray diffraction revealed that it crystallizes in the orthorhombic TiNiSi-type structure with the $Pnma$ space group. 
The structure comprises alternately stacked layers of Mg and PtSi honeycomb network, which is reminiscent of MgB$_2$, and the buckling of the honeycomb network causes orthorhombic distortion.
Electrical and magnetic studies revealed that MgPtSi exhibited superconductivity with a transition temperature of 2.5 K.
However, its isostructural compounds, namely, MgRhSi and MgIrSi, were not found to exhibit superconductivity.
\end{abstract}



\maketitle

\section{Introduction}
Compounds with honeycomb structures and their derivatives often exhibit intriguing superconductivity, which partly stems from the lack of spatial inversion symmetry and the presence of strong spin-orbit coupling.
For example,
chiral $d$-wave superconductivity\cite{Fischer}, triplet $f$-wave superconductivity,\cite{Wang} and their combination\cite{Goryo} have been theoretically predicted to be characteristic of SrPtAs.
This prediction is the subject of ongoing experimental confirmation.\cite{Biswas,Matano} 
The superconductivity of Ru$_{1-x}$Rh$_x$P emerges around the critical point of the pseudo-gap phase on the verge of the metal-insulator transition,\cite{Hirai1} while that of CrAs emerges at the helimagnetic quantum critical point.\cite{Wu,Kotegawa} 
Furthermore, the coexistence of superconductivity and a ferromagnetic order, as well as magnetic-field-reentrant superconductivity, has been observed in UCoGe\cite{Huy} and URhGe.\cite{Aoki,Aoki2}

The honeycomb structure is best exemplified by MgB$_2$, which crystallizes in the AlB$_2$-type structure ($P6/mmm$, No.~191) and is a superconductor with a transition temperature $T_{\rm c}$ of 39 K.\cite{Nagamatsu} 
As an ordered variant of the AlB$_2$-type structure, the Ni$_2$In-type structure ($P6_{3}/mmc$, No. 194) is formed when Ni and In atoms are substituted for B to form ordered NiIn honeycomb layers, and the remaining Ni atoms are substituted for the Al atoms positioned between them. 
The replacement of the interlayer Ni in this structure with another element results in the KZnAs-type structure ($P6_3/mmc$, No. 194). 
This crystal structure is found in the aforementioned SrPtAs, which has a $T_{\rm c}$ of 2.4 K.\cite{Nishikubo} 
Another ordered variant of the AlB$_2$-type structure is the SrPtSb-type structure ($P\bar{6}m2$, No.~187). 
$1H$-CaAlSi, which exhibits this latter structure, is characterized by superconductivity at a relatively high $T_{\rm c}$ of 6.5 K.\cite{Imai,Kuroiwa} 
Furthermore, the removal of the in-plane Ni atoms in the above-mentioned Ni$_2$In-type structure produces the NiAs-type structure ($P6_3/mmc$, No.~194), from which the MnP-type structure ($Pnma$, No.~62) can then be obtained by orthorhombic distortion. 
IrGe ($T_{\rm c}$ = 4.7 K),\cite{Hirai2} the aforementioned Ru$_{1-x}$Rh$_x$P ($T_{\rm c}$ = 3.7 K),\cite{Hirai1} and CrAs ($T_{\rm c}$ = 2.2 K under hydrostatic pressure)\cite{Kotegawa,Wu} crystallize in the MnP-type structure. 
The TiNiSi-type structure ($Pnma$, No.~62) is another orthorhombic variant of the AlB$_2$-type structure that is derived from the distortion of the Ni$_2$In-type structure. 
Examples of compounds with the TiNiSi-type structure are Mg(Mg$_{1-x}$Al$_x$)Si ($T_{\rm c}$ $\sim$ 6 K),\cite{Ji} the aforementioned UCoGe ($T_{\rm c}$ = 0.8 K),\cite{Huy} and URhGe ($T_{\rm c}$ = 0.25 K),\cite{Aoki} the distorted honeycomb layers of which are composed of (Mg$_{1-x}$Al$_x$)Si, CoGe, and RhGe, respectively. 
Thus, the derivatives of the AlB$_2$-type structure afford researchers the opportunity to explore novel superconducting materials.

In this paper, we report the superconductivity of the compound MgPtSi. 
Powder X-ray diffraction (XRD) analysis conducted on the synthesized compound revealed that it crystallizes in the orthorhombic TiNiSi-type structure ($Pnma$ space group), as shown in Fig.~1. 
This structure consists of buckled honeycomb layers of PtSi alternated with layers of Mg to form an orthorhombic variant of MgB$_2$, which has regular honeycomb layers composed of boron atoms. 
Electrical and magnetic measurements revealed the superconductivity of MgPtSi at a $T_{\rm c}$ of 2.5 K. 
\begin{figure}[t]
\begin{center}
\includegraphics[width=6.5cm]{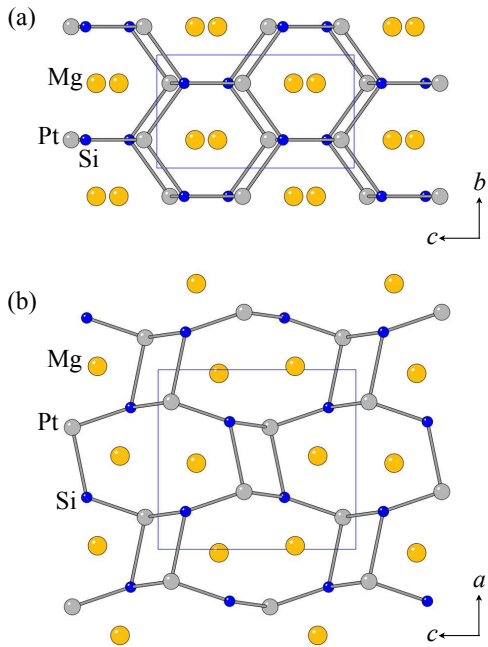}
\caption{(color online) Crystal structure of TiNiSi-type MgPtSi with the $Pnma$ space group (No. 62) viewed along the (a) $a$-axis and (b) $b$-axis. The solid lines indicate the unit cell.}
\label{f1}
\end{center}
\end{figure}

\section{Experimental}
Polycrystalline samples of MgPtSi were synthesized by heating a mixture of Mg$_2$Si, Pt, and Si powders with a ratio of Mg:Pt:Si of 10:7:25. 
The mixture was placed in an alumina crucible and sealed in a quartz tube under partial Ar pressure to minimize Mg evaporation during the reaction. 
The samples were heated at 1000$^\circ$C for 12 h, and this was followed by furnace cooling, which yielded MgPtSi together with a significant amount of Si.  No other impurities were present. 
The use of other synthesis conditions produced several impurities such as Pt, PtSi, Mg$_2$Si, Mg$_2$Pt, Mg$_2$PtSi, Mg$_5$Pt$_{10}$Si$_{16}$, and other unidentified substances. 
None of these impurities exhibited superconductivity down to 1.8 K. 
The structures of the MgPtSi samples were examined by powder XRD using a Rigaku RINT-TTR III X-ray diffractometer with Cu$K_{\alpha}$ radiation.  Rietveld refinement was performed using the RIETAN-FP program,\cite{Izumi} and the magnetization $M$ was measured using the Quantum Design Magnetic Property Measurement System (MPMS). The electrical resistivity $\rho$ was measured using the Quantum Design Physical Property Measurement System (PPMS).
\begin{table}[h]
\caption{
Rietveld-refined MgPtSi structural parameters at 300 K, for the orthorhombic structure with the $Pnma$ space group (No. 62). 
$a$ = 6.5324(3) \AA, $b$ = 4.1005(2) \AA, and c = 7.1752(4) \AA.
Occupancy, atomic coordination, and isotropic atomic displacement parameter are listed. The weighted residual of the least-squares refinement $R_{wp}$ is 9.826\%, and the residual of the least-squares refinement $R_{p}$ is 7.512\%. The occupancy $g$ is fixed.
}
\begin{ruledtabular}
\begin{tabular}{ccccccc}
site & atom & $g$ & $x/a$ & $y/b$ & $z/c$ & $100U$ (\AA$^2$)\\
\hline
4c & Mg & 1   &  -0.020(2)  &  1/4 &  0.693(2) & 0.3(4)\\
4c & Pt & 1   &   0.1788(3)  & 1/4  & 0.0658(3) & 0.8(2)\\
4c & Si & 1     &  0.289(2)     & 1/4  &  0.361(2)  & 1.2(4)\\
\end{tabular}
\end{ruledtabular}
\end{table}

\section{Results and Discussion}
\subsection{Structure}
As shown in Fig.~2, it was possible to refine the powder XRD profile of the obtained sample based on the orthorhombic TiNiSi-type structure with the $Pnma$ space group (No. 62). 
The observed peaks, which had intensities as high as 10$^5$ counts, clearly confirm the TiNiSi-type structure, as can be seen from Fig.~2(b), although the several peaks are due to the elemental Si, which has the $Fm\bar{3}m$ space group (No. 225). 
The weight ratio of MgPtSi:Si was estimated to be 15:85 for the sample used in the structural analysis. 
The obtained crystallographic data of the MgPtSi are summarized in Table I. 
Partial substitution of Mg in the Pt sites, as was observed in the case of Mg(Al$_{1-x}$Mg$_x$)Si,\cite{Ji} was not found in the present compound.
\begin{figure}[t]
\begin{center}
\includegraphics[width=8cm]{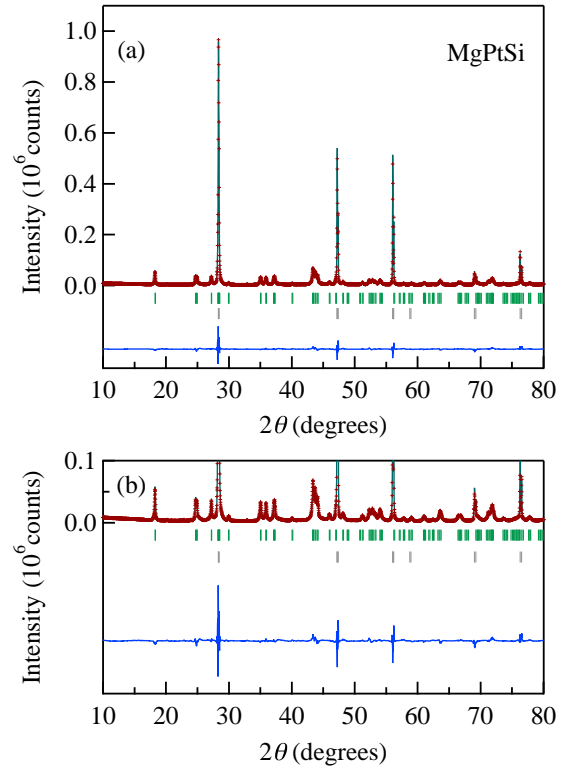}
\caption{(color online) (a) Powder XRD of MgPtSi obtained at room temperature, and its Rietveld refinement. (b) Enlarged view showing peaks with intensities of up to $0.1 \times 10^6$ counts. The red circles, black line, and blue line indicate the observed, calculated, and difference profiles, respectively. The green and gray ticks respectively indicate the calculated Bragg diffraction positions of the MgPtSi with the $Pnma$ space group and Si with $Fm\bar{3}m$. These two phases are taken into account in the calculated profile.}
\label{f1}
\end{center}
\end{figure}

From the crystal structure of MgPtSi shown in Fig.~1, it can be seen that Pt and Si alternately occupy sites in the honeycomb network. 
This is apparent from the $b$-$c$ projection of the structure shown in Fig.~1(a). 
Mg is located between the honeycomb layers. 
In addition, the honeycomb layers exhibit buckling along the crystallographic $c$-axis, and this causes orthorhombic distortion. 
The buckling of adjacent PtSi layers occurs in opposite directions along the $a$-axis, resulting in the formation of interlayer chemical bonds between Pt and Si along this axis, as is apparent from the $a$-$c$ projection of the crystal structure shown in Fig.~1(b). 
The length (2.597 {\AA}) of the Pt-Si bond between the honeycomb layers is comparable to those within the honeycomb layers (2.241 and 2.529 {\AA}), suggesting that the structure can be viewed as a three-dimensional network composed of Pt and Si, rather than a two-dimensional honeycomb network.

Three types of buckling are known to occur in the honeycomb network with the TiNiSi-type structure.\cite{Landrum} 
They are ``chair"-type buckling, in which one atom in the honeycomb hexagon is displaced upward while the atom positioned on the opposite side of the hexagon is displaced downward; ``boat"-type buckling, in which both atoms are displaced in the same direction; and ``half-chair"-type buckling, in which only one atom is displaced. 
The buckling that occurs in MgPtSi is of the ``chair"-type, as can be observed from Fig.~1(b), while Mg(Mg$_{1-x}$Al$_x$)Si exhibits ``boat"-type buckling.\cite{Ji}

\subsection{Superconductivity}
Figure 3(a) shows the temperature dependence of $\rho$ for MgPtSi under zero-field conditions. 
The resistivity was metallic in character and dropped sharply below 2.97 K, as is typical of transition to superconductivity. 
Zero resistivity was observed at 2.52 K, and the 10\%--90\% transition width was estimated to be 0.20 K, as shown in Fig.~3(b). 
Bulk superconductivity in MgPtSi is evidenced by the temperature dependence of $M$ shown in Fig.~4. 
It exhibits a diamagnetic behavior below 2.5 K. 
Furthermore, the shielding and flux exclusion signals correspond to 50\% and 86\% perfect diamagnetism, respectively. 
These observations have led to our conclusion that MgPtSi is a superconductor with $T_{\rm c}$ = 2.5 K. 
\begin{figure}[t]
\begin{center}
\includegraphics[width=8cm]{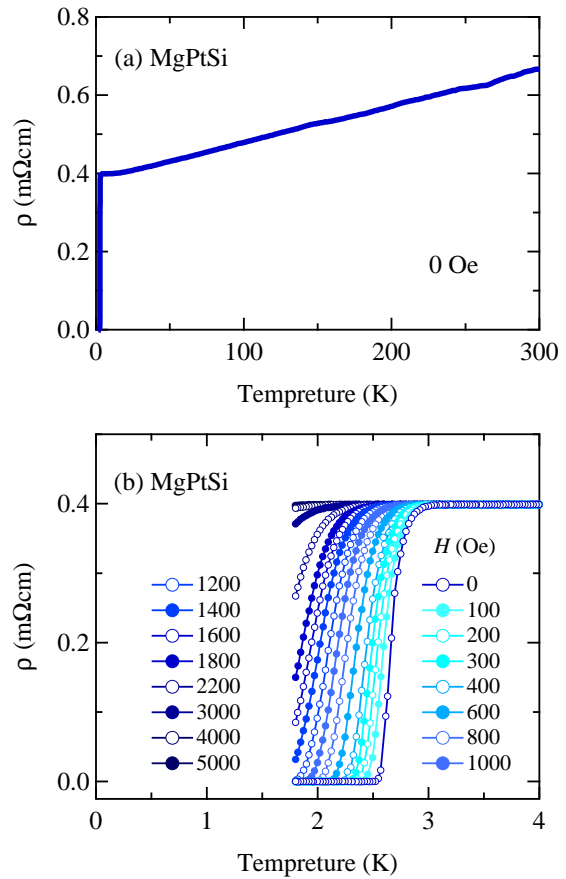}
\caption{(color online) Temperature dependence of electrical resistivity $\rho$ of MgPtSi in (a) zero field and (b) magnetic fields of up to 5000 Oe.}
\label{f1}
\end{center}
\end{figure}

The $T_{\rm c}$ value for MgPtSi decreases with increasing magnetic field and disappears at 5000 Oe, as shown in Fig.~3(b). 
The magnetic field dependence of $\rho$ indicates that MgPtSi is a type-II superconductor. 
Figure 5 shows the temperature dependence of the upper critical field $H_{c2}$, which was determined from the midpoint of the resistive transition. 
It can be seen that, at low temperatures below 2.35 K, $H_{\rm c2}$ increases almost linearly with decreasing temperature, down to 1.86 K. 
By linear extrapolation, we estimated $H_{\rm c2}$ at 0 K [i.e., $H_{\rm c2}$(0)] to be 6360 Oe. 
The Ginzburg-Landau coherence length $\xi(0)$ was also estimated to be 230 \AA \ based on the relationship $\xi(0) = [\Phi_0/2\pi H_{\rm c2}(0)]^{1/2}$, where $\Phi_0$ is the magnetic flux quantum. 
It is worth noting here that close inspection near $T_{\rm c}$ revealed an upward curvature of $H_{\rm c2}(T)$, suggesting multiband/multigap superconductivity of the present compound. 
A similar but more prominent behavior has been reported for MgB$_2$.\cite{Gurevich1,Gurevich2}

The emergence of superconductivity in MgPtSi may be related to the occurrence of structural instability. 
The superconductivity of many compounds with the AlB$_2$-type structure has been found to emerge close to the structural phase transition. 
Examples of such compounds are Sr(Al,Si)$_2$,\cite{Lorenz,Evans} Sr(Ga,Si)$_2$,\cite{Imai2,Meng} SrNi$_x$Si$_{2-x}$,\cite{Pyon} and BaTM$_x$Si$_{2-x}$ (TM = Pd, Pt, Cu, Ag, and Au).\cite{Kawashima} 
These compounds are on the verge of structural phase transition to the cubic SrSi$_2$-type structure ($P4_{1}32$, No. 212), and their superconducting AlB$_2$-type structure is stabilized by partial chemical substitutions for Si. 
Analogous structural instability toward a cubic phase can be inferred for the present MgPtSi based on the consideration of the structural transition of CaPtSi under high pressure.\cite{Evers} 
It should be noted that Mg and Ca are both isoelectronic elements, which justifies the following comparison. 
CaPtSi crystallizes in the cubic LaIrSi-type structure ($P2_13$, No.~198), which is a derivative of the SrSi$_2$-type structure, and transforms to the orthorhombic TiNiSi-type structure at high pressures of 3--4 GPa.\cite{Evers} 
Because the ionic radius of Mg$^{2+}$ (0.72 {\AA}) is significantly smaller than that of Ca$^{2+}$ (1.00 {\AA}), the substitution of Mg for Ca results in the generation of a large chemical pressure, and MgPtSi can thus be regarded as a high-pressure analog of CaPtSi. 
In other words,
the TiNiSi-type structure of MgPtSi becomes unstable on the verge of transition to the cubic-phase LaIrSi-type structure.
We thus predict that the partial substitution of Ca for Mg would amplify this structural instability and thereby increase the superconducting transition temperature of MgPtSi. 
However, the results of our preliminary experiments suggested that Ca doping actually decreased the $T_{\rm c}$ to 2.2 K. 
This might have been due to the disorder caused by the large difference between the ionic radii of Mg and Ca. 
\begin{figure}[t]
\begin{center}
\includegraphics[width=8cm]{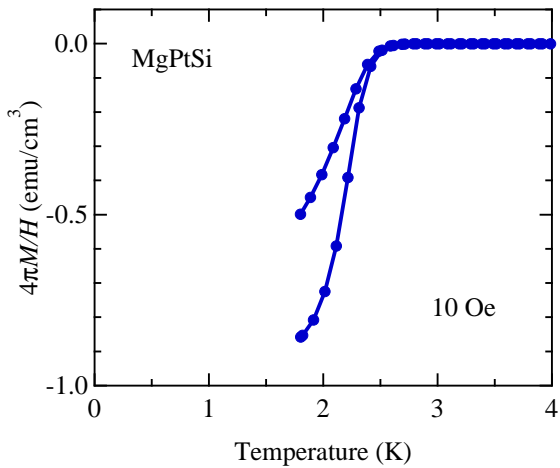}
\caption{(color online) Temperature dependence of magnetization divided by magnetic field, $M/H$, for MgPtSi in a magnetic field of 10 Oe under zero-field and field cooling conditions. }
\label{f1}
\end{center}
\end{figure}
\begin{figure}[t]
\begin{center}
\includegraphics[width=8cm]{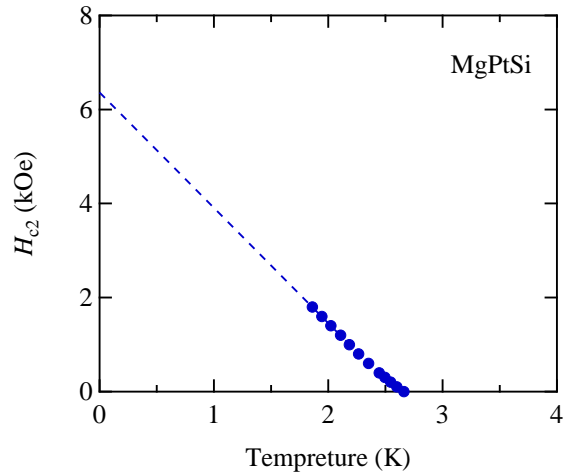}
\caption{(color online) Temperature dependence of upper critical field $H_{\rm c2}$ of MgPtSi determined by resistivity measurements. The broken line represents the linear extrapolation.}
\label{f1}
\end{center}
\end{figure}

Another means of increasing $T_{\rm c}$ is by controlling the number of charge carriers. 
We implemented this procedure by substituting Rh and Ir for Pt to reduce the number of valence electrons by one and successfully obtained two isostructural compounds, namely, MgRhSi and MgIrSi. 
In this case, the lattice parameters were $a$ = 6.5719(2) and 6.5740(3) {\AA}, $b$ = 4.0110(1) and 4.0149(2) {\AA}, and $c$ = 6.9798(3) and 6.9693(4) {\AA}, for MgRhSi and MgIrSi, respectively.
These compounds did not exhibit superconductivity, at least down to 1.8 K. 
In contrast, the partial substitution of Au for Pt to increase the number of valence electrons resulted in an increase of the $T_{\rm c}$ of Mg(Pt$_{1-x}$Au$_x$)Si to 2.9 K with a nominal $x$ value of 0.5. 
These results cannot be understood in terms of the change in the electronic density of states (DOS) at the Fermi level ($E_{\rm F}$) after chemical doping, which invokes a change in the Fermi surface topology or nesting, as discussed below.

\subsection{Electronic band structure}
In this section, we present the results of first-principles calculations for MgPtSi obtained by the WIEN2k package.\cite{Blaha} 
The structural parameters in Table I were used for the calculations. 
An examination of the electronic DOS in Fig.~6 reveals a strong Pt-$5d$ character at $E_{\rm F}$ and almost the same amounts of interstitials, while the contributions of Mg and Si are very small. 
This contrasts markedly with the case of MgB$_2$, in which the B-$2p$ orbitals dominate the Fermi level together with the interstitials,\cite{kortus} and the charge carriers in the strongly covalent B-$2p$-$\sigma$ bands enhance the coupling to high-energy optical phonons, resulting in a relatively high $T_{\rm c}$ of 39 K.\cite{choi} 
In MgAlSi, which has a buckled honeycomb layer of AlSi, both the $3s$ and $3p$ orbitals of Al and Si contribute to a relatively large DOS of approximately 5 states/eV at $E_{\rm F}$, resulting in a $T_{\rm c}$ of $\simeq$ 6 K.\cite{Ji} 
In the present MgPtSi, a small DOS of 3.1 states/eV with a strong Pt-$5d$ character, which produces a weak electron-phonon coupling, is the most probable reason for the lower $T_{\rm c}$ of  2.5 K. 
\begin{figure}[t]
\begin{center}
\includegraphics[width=7.8cm]{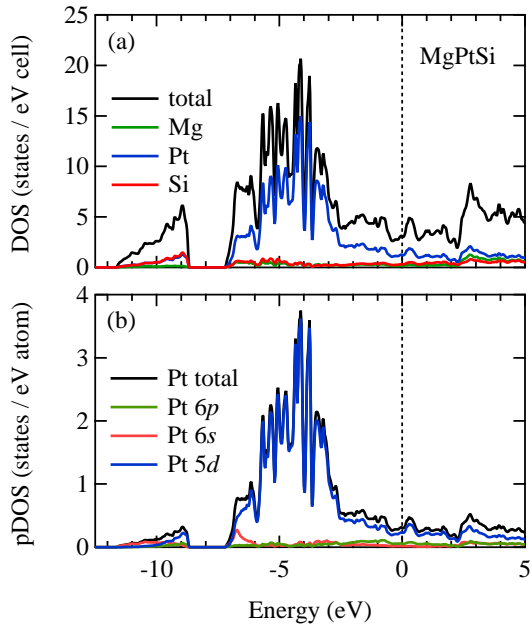}
\caption{(color online) (a) Total and (b) partial electronic density of states of MgPtSi. Fermi level $E_{\rm F}$ = 0 eV. 
The spin-orbit (SO) interaction has not been taken into consideration. 
We have checked that little modification results from considering the SO interaction of Pt.
}
\label{f1}
\end{center}
\end{figure}
\begin{figure}[h]
\begin{center}
\includegraphics[width=7.8cm]{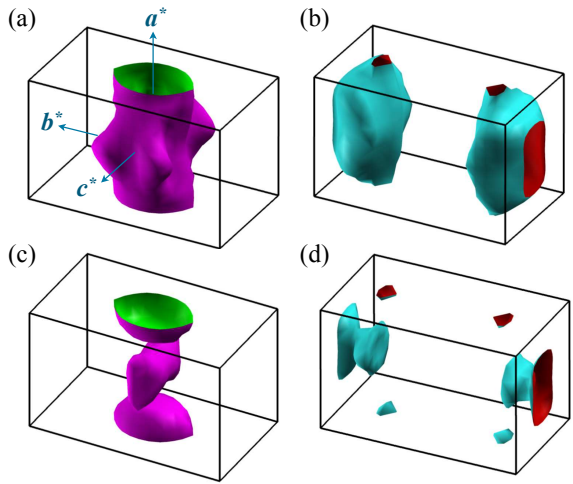}
\caption{(color online) Fermi surface of MgPtSi obtained by the WIEN2k package. 
}
\label{f1}
\end{center}
\end{figure}

Although the Pt--Si network is fairly three-dimensional owing to the Pt--Si chemical bonds formed between the buckled PtSi honeycomb layers, an essentially two-dimensional Fermi surface exists in MgPtSi. 
This suggests that the PtSi honeycomb layers can be considered as the primary conducting layers. 
The Fermi surface is almost cylindrical at the $\Gamma$ point, as shown in Fig.~7(a), and contains several three-dimensional shapes, as shown in Figs. 7(b)--(d). 
This is reminiscent of the Fermi surface of MgB$_2$, which is cylindrical at the $\Gamma$ point and contains a three-dimensional tabular network.\cite{kortus} 
However, the strong Pt-$5d$ character of MgPtSi produces a low $T_{\rm c}$.  
It is our expectation that if the Pt--Si bonds between the honeycomb layers of MgPtSi are broken, the Fermi surface would become more two-dimensional and the $T_{\rm c}$ would increase. 
Nevertheless, the $T_{\rm c}$ would be much lower than that of MgB$_2$ because of the weaker electron-phonon coupling of MgPtSi.

Finally, we argue the non-superconductivity of MgRhSi and MgIrSi. 
With the assumption of a rigid band,\cite{structure} we estimate that MgRhSi and MgIrSi have an $E_{\rm F}$ that is approximately 1.1 eV lower than that of MgPtSi, and that their DOSs are approximately 4.9 states/eV. 
Relatively higher DOSs of MgRhSi and MgIrSi compared to MgPtSi would also cause their $T_{\rm c}$ values to be higher. 
The absence of superconductivity in MgRhSi and MgIrSi is thus significant, and the topological modification of the Fermi surfaces of these compounds by Rh (Ir) substitution for Pt may be used to reverse the condition.

\section{Conclusions}
A ternary compound, MgPtSi, was developed by conventional solid-state reaction, and electrical and magnetic measurements revealed its superconductivity at a relatively low transition temperature $T_{\rm c}$ of 2.5 K. 
The crystal structure of the compound consists of alternately stacked PtSi honeycomb layers and Mg layers, and the buckling of the honeycomb layers induces the formation of Pt-Si chemical bonds between them. 
The formation of these bonds makes MgPtSi into a three-dimensional analog of MgB$_2$, which has a $T_{\rm c}$ of 39 K. 
There are a large number of compounds that have the AlB$_2$-type structural class, and this makes the structure worthy of further exploration with the purpose of discovering and investigating the properties of new superconducting materials.

\section{Acknowledgments}
This work was partially supported by Grants-in-Aid for Scientific Research from the Japan Society for the Promotion of Science (JSPS) (Grants No. 25400372 and No. 26287082).


\begin{thebibliography}{99}
\bibitem{structure} The atomic positions of MgRhSi and MgIrSi were not reliably refined in the XRD profiles because of the obstructive peaks of the impurity phases. Thus we used a ridged band picture to estimate the DOSs of MgRhSi and MgIrSi.
\end{thebibliography}
\end{document}